%% file: main.tex
\documentclass[journal]{vgtc}                     

\vgtccategory{Research}

\newcommand{\alexin}[1]{{\textcolor[HTML]{000000}{#1}}}
\newcommand{\tool}{\textit{ChartifyText}} 

\title{\tool{}: Automated Chart Generation from Data-Involved Texts via LLM}

\author{%
Songheng~Zhang, Lei~Wang, Toby~Jia-Jun~Li, Qiaomu~Shen, Yixin~Cao and Yong~Wang}

\abstract{%
Text documents with numerical values involved are widely used in various applications such as scientific research, economy, public health and journalism. However, it is difficult for readers to quickly interpret such data-involved texts and gain deep insights.
To fill this research gap, this work aims to automatically generate charts to accurately convey the underlying data and ideas to readers, which is essentially a challenging task. The challenges originate from text ambiguities, intrinsic sparsity and uncertainty of data in text documents, and subjective sentiment differences. 
Specifically, we propose \tool{}, a novel fully-automated approach that leverages Large Language Models (LLMs) to convert complex data-involved texts to expressive charts. It consists of two major modules: tabular data inference and expressive chart generation.
The tabular data inference module employs systematic prompt engineering to guide the LLM (e.g., GPT-4) to infer table data, where data ranges, uncertainties, missing data values and corresponding subjective sentiments are explicitly considered. The expressive chart generation module augments standard charts with intuitive visual encodings and concise texts to accurately convey the underlying data and insights.
We extensively evaluate the effectiveness of \tool{} on real-world data-involved text documents through case studies, in-depth interviews with three visualization experts, and a carefully-designed user study with 15 participants.
The results demonstrate the usefulness and effectiveness of \tool{} in helping readers efficiently and effectively make sense of data-involved texts.

}

\vspace{-2em}
\teaser{
  \centering
  \includegraphics[width=0.96\linewidth]{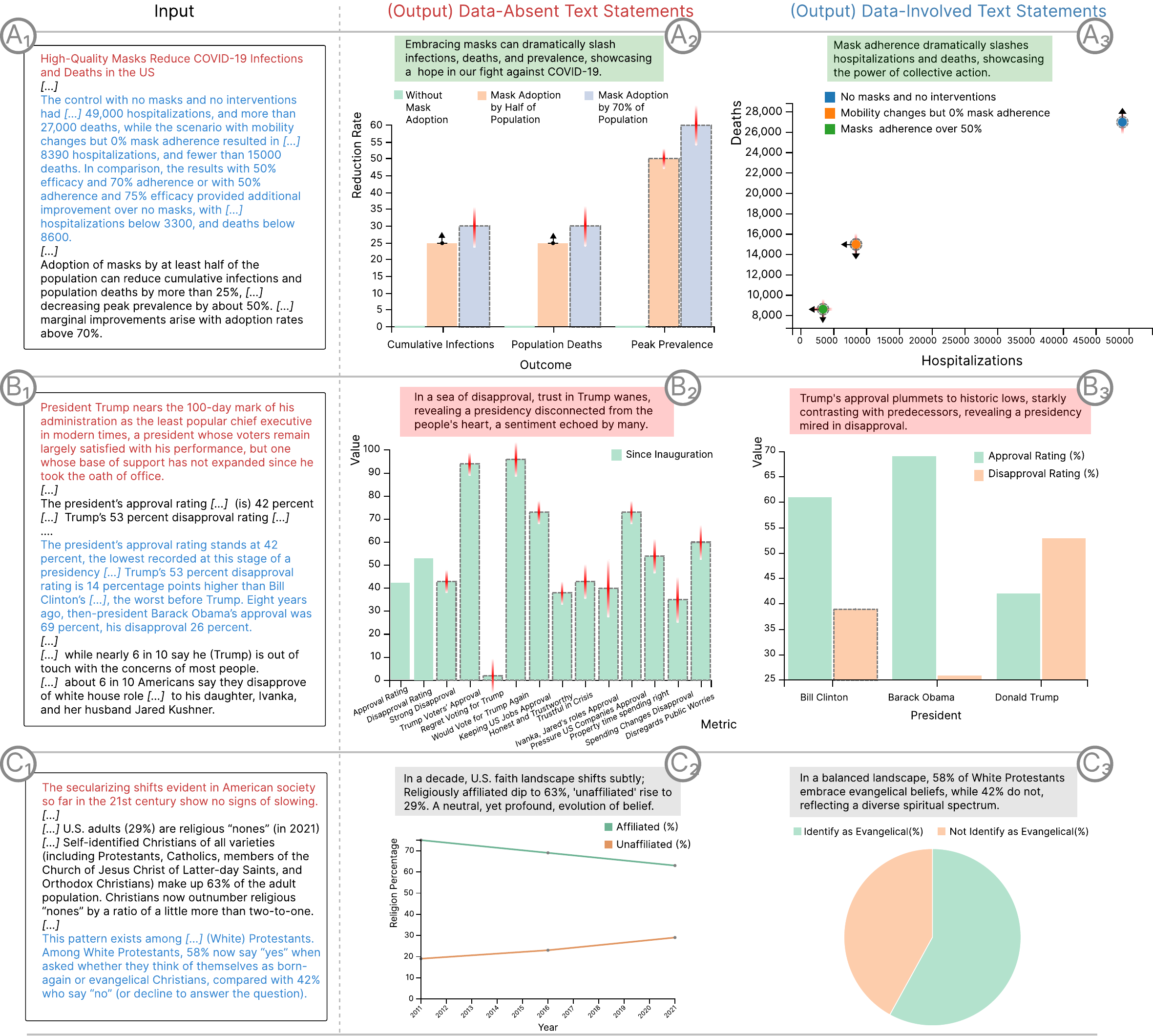}

      \caption{%
  \tool{} generates intuitive charts based on user-selected text statements. The first column displays document excerpts. \textcolor{lightred}{Data-absent text statements} do not contain any data. \textcolor{lightblue}{Data-involved text statements} contain data. These statements' contexts are shown in black. The right two columns showcase charts generated from two types of statements. These charts visualize the text statements. The middle column showcases charts generated from \textcolor{lightred}{data-absent text statements}, and the right column showcases charts generated from \textcolor{lightblue}{data-involved text statements}. The charts' text annotations link the statements to the data points. The background color of the text annotation represents the sentiment in the statements: light green for positive, light red for negative, and light gray for neutral.}
  \label{fig:teaser}
}
\graphicspath{{figs/}{figures/}{pictures/}{images/}{./}} 

\usepackage{tabu}                      
\usepackage{booktabs}                  
\usepackage{lipsum}                    
\usepackage{mwe}                       

\usepackage{mathptmx}                  
\usepackage{graphicx}
\usepackage{multirow}
\usepackage{multicol}
\usepackage{tabularray}
\usepackage{tabularray}
\usepackage{ulem}

\newcommand\st{\bgroup\markoverwith{\textcolor{black}{\rule[0.5ex]{2pt}{0.4pt}}}\ULon}

\begin{document}


\definecolor{lightblue}{RGB}{35, 132, 221}
\definecolor{lightred}{RGB}{203, 71, 71}

\maketitle

\input{src/1-intro}

\input{src/2-relatedwork}

\input{src/3-method}

\input{src/4-evaluation}

\input{src/5-discussion}
\input{src/6-conclusion}
\bibliographystyle{abbrv-doi-hyperref}

\bibliography{main}
\clearpage

\input{src/7-appendix}
\end{document}

%% file: src/1-intro.tex
\section{Introduction}





    


Text documents are often filled with numerical values to report data analysis findings or justify specific claims.
With the world being increasingly data-driven, such kind of \textbf{\textit{data-involved texts}} can be easily seen in every application domain that uses data, such as scientific research, economy, public health and journalism.
Despite the popularity and importance of data-involved texts, it is difficult to quickly gain deep insights into the involved numerical values~\cite{Masson2023CharagraphIG}, due to the intrinsic sparse distribution of numerical values and the linear structure of texts.
For example, for the highlighted blue texts in Figure~\ref{fig:teaser} B\textsubscript{1}, if we want to know which president has the highest disapproval rating, we need to identify and remember all the relevant percentage numbers from the whole text document, mentally calculate the disapproval rating of \textit{Bill Clinton}, and further mentally compare the disapproval ratings of all three presidents.
There is a high cognitive load in such a process, resulting in the inefficiency of digesting data-involved texts.
Due to this reason, prior research has also criticized the practice of using texts to present numbers~\cite{tufte2001visual,feliciano1963presentation,klein2014communicating}, despite the wide usage of data-involved texts.

Meanwhile, data visualization has demonstrated its great power in significantly augmenting human users' capability of perceiving data from various domains~\cite{munzner2014visualization,van2005value}, and it is widely accepted that ``one chart is worth ten thousand words''~\cite{larkin1987diagram}.
At a quick glance, viewers can quickly understand the main idea from charts.
Inspired by this, we pose one crucial research question: \textit{can we \textbf{\underline{automatically} generate expressive charts from data-involved texts} to \textbf{\underline{accurately}} convey their underlying data and insights?} 
It is a straightforward question to ask, but a very challenging task to achieve.
There are challenges for both \textit{\textbf{data extraction}} from data-involved texts and \textbf{\textit{chart generation}} from such extracted data.

For data extraction, the challenges originate from the essential complexity of data-involved texts, including \textit{text ambiguity}, \textit{data sparsity} and \textit{subjective sentiment difference}.
First, text descriptions can be ambiguous. For example, an article talking about the mask usage during COVID 19 may refer to United States by using different names like \textit{``United States''}, \textit{``US''}, \textit{``USA''} and \textit{``America''}. It can also use ambiguous or uncertain words to delineate the numerical values like \textit{``more than 27,000 deaths''}, \textit{``below 3300''} and \textit{``about 50\%''}, as shown in Figure~\ref{fig:teaser} A\textsubscript{1}.
It leads to \textit{\textbf{data ranges}} and associated \textit{\textbf{uncertainties}} for data ranges or points, instead of the generally-expected concrete data points in data visualization.
Second, the numerical values are intrinsically sparse in texts, since it is rarely seen that all the numbers of a dataset are presented as texts in a document, and only a few numbers are sparsely and linearly distributed in a text document.
It can result in inevitable \textit{\textbf{missing data values}}, which may or may not be able to be inferred from the text document.
Third, different from pure data, texts often incorporate the authors' \textit{\textbf{subjective sentiments}} or viewpoint~\cite{medhat2014sentiment,wankhade2022survey}. For the same data values, different authors may use totally different sentiments to introduce them. Thus, it is also crucial to accurately extract and reflect the subjective sentiment in data-involved texts.

For chart generation, the challenges come from \textit{\textbf{accurately}} visualizing the data to be extracted from data-involved texts.
%
%
%
%
%
There has been extensive research on automated visualization recommendation and generation for a common and concrete dataset~\cite{wang2021survey, zhang2023adavis,zhou2021table2charts}. But none of them has coped with the data extracted from data-involved texts, which intrinsically incorporates the issues of data ranges, uncertainties, missing data values and subjective sentiment. It is non-trivial to accurately visualize such data in a systematic manner. Three recent research has also attempted to create visualizations from texts~\cite{Masson2023CharagraphIG,cui2019text, Rashid2021Text2ChartAM}. 
However, these approaches have limitations. They depend on manual input from users, necessitate substantial training data, or are restricted to specific data types, such as percentages, limiting their applicability for automated chart generation.
Also, None addresses the above challenges intrinsic to data extracted from data-involved texts.

In this work, we fill the research gap by proposing \tool{}, a novel fully-automated approach to transform data-involved texts into intuitive charts. It is built upon LLMs and systematically tackles the above challenges in data extraction and chart generation from data-involved texts.
Specifically, \tool{} takes the whole text document and specific sentences of users' interest as input, and consists of two major modules: 
\textit{tabular data inference} and \textit{expressive chart generation}.
The tabular data inference module leverages GPT-4~\cite{Achiam2023GPT4TR} to analyze the sentences specified by users to identify their possible data exploration topics and formulate the table schema. Then by designing appropriate prompts, the tabular data inference module further infers the individual data values for each cell in the table by explicitly considering data ranges, inference uncertainties, missing data values and subjective sentiment.
With the tabular data inferred from the data-involved texts, we propose intuitive and concise visual encodings to augment standard charts (e.g., bar chart, pie chart, line chart and scatter plot), accurately conveying the underlying data with data ranges, uncertainties, missing values and different subjective sentiments to viewers.
By viewing the augmented charts automatically generated by \tool{}, users can quickly and accurately understand the underlying data and data insights of their interest. 
\alexin{We evaluated the usability and effectiveness of \tool{} through expert interviews and a user study. The expert interviews focused on the informativeness and accuracy of the charts generated by \tool{}, assessing their ability to represent data-involved texts effectively. Meanwhile, the user study explored whether \tool{} enabled users to quickly and accurately understand data-involved texts.}

In summary, the contributions of this work can be summarized as follows:

\begin{itemize}
    \item  We propose \tool{}, a novel approach to automatically transform data-involved texts into intuitive and expressive visualizations, facilitating an efficient and accurate comprehension of data-involved texts. It consists of a tabular data inference module and an expressive chart generation module, which is built upon LLMs and explicitly addressed the challenges inherent in data-involved texts.

    
    
    \item We conduct extensive evaluations on real-world data-involved text documents, including case studies, expert interviews and user studies, to comprehensively assess \tool{}. The results demonstrate the effectiveness and usability of \tool{} in allowing users easily gain insights into data-involved text documents.

    
\end{itemize}

%% file: src/2-relatedwork.tex
\section{Related work}
The related work of this paper falls into three categories: automatic visualization generation from text, data table extraction from text and data visualization with missing items.

\subsection{Automatic Visualization Generation from Text}
The concept of automatically providing data visualizations for text has seen considerable advancement. The concept aims to improve reader comprehension of articles by supplementing them with relevant visual aids. Existing methods fall into two categories: retrieval-based and generation-based. Retrieval-based methods extract visualizations related to a particular text content from an existing source. On the other hand, generating-based methods are capable of producing visualizations customized to the text content.
Retrieval-based methods, exemplified by VizByWiki~\cite{Lin2018VizByWikiMD}, retrieved visualizations from existing web content, tailored to news articles. However, this approach is hampered by the reliance on external visualization quality and datedness, often leading to a mismatch with current articles.
Generation-based methods, like Conexifier~\cite{Hullman2013ContextifierAG} and NewsViews~\cite{Gao2014NewsViewsAA}, create context-specific visualizations such as stock price line charts or interactive maps. These, however, necessitate external datasets and are confined to particular visualization types.
To address these constraints, Masson~\textit{et al.} introduced Charagraph~\cite{Masson2023CharagraphIG} and Statslator~\cite{Masson2023StatslatorIT}, capable of generating interactive visualizations from in-text numerical data in real time, without external data dependencies. \alexin{Additionally, Rashid \textit{et al.} leverages the LSTM to generate chart from text in an end-to-end way.} Yet, their application is limited to explicit numerical data within the text, overlooking the potential of non-numeric information.

Our proposed solution, \tool{} harnesses the inference capabilities of Large Language Models (LLMs). It adeptly infers numerical values from non-numeric text data. For instance, \tool{} can discern and extract values such as ``3'' and ``6'' from a sentence like ``Alex has 3 apples, and Bob's apple number is twice Alex's''.

\subsection{Data Table Extraction from Text}
The topic of text-to-table generation, which aims to transform textual information into structured tabular formats without predefined schemas such as table headers, has seen significant advancements in recent years. 
Wu \textit{et al.}~\cite{Wu2021TexttoTableAN} laid the foundational groundwork in this area by developing a data-driven sequence-to-sequence (Seq2Seq) approach, which converts text into tabular data. Building on this, Li \textit{et al.}~\cite{Li2023ASM} separates the generation of table headers and bodies, thereby improving the model's accuracy of the text-to-table tasks.
the STable~\cite{Pietruszka2022STableTG} framework leverages sequence-to-sequence models for converting text into structured tables as well. However, unlike previous works, the model will explicitly predict of the number of rows before the table generation starts. Furthermore, Li \textit{et al.}~\cite{Li2022TowardAU} utilized unsupervised learning techniques to generate synthetic text-table pairs without human annotations. With the synthetic text-table pairs. This method can significantly reduce the dependence on manual annotations.
However, their models do not address user-specific needs. Jiao \textit{et al.}'s~\cite{Jiao2023InstructAE} introduction of Large Language Models (LLMs) marked a shift towards more user-centric text-to-table conversions, enabling users to specify table headers for more relevant information extraction. 

Our work builds upon Jiao \textit{et al.}'s foundation~\cite{Jiao2023InstructAE} but goes a step further. It not only extracts table content from the text but also converts non-numeric information into numeric data, such as specific figures or value ranges with \alexin{uncertainty} scores, enhancing the depth and utility of the extracted information. Unique to our approach is the LLMs' ability to provide \alexin{uncertainty} scores for these conversions, offering an additional layer of reliability. As a result, the converted tables are readily usable in data visualization.

\subsection{Visualization of Missing Data}
Missing data is a frequent challenge in data analysis, and its visualization plays a pivotal role in comprehensive data interpretation~\cite{Song2019WheresMD}. To address this, two primary strategies are employed~\cite{Song2021UnderstandingTE, Sarma2022EvaluatingTU}. First, the visualization of the missing data itself is exemplified by tools such as xGobi~\cite{Swayne1998MissingDI}, MANET~\cite{Unwin1996InteractiveGF}, and VIM~\cite{Templ2011ExploringID}~\alexin{, which allows analysts to directly notice missing values. Song~\textit{et al.} further investigates various techniques like color coding to emphasize (I.e., highlighting) or minimize (i.e., downplaying) the visibility of missing data points~\cite{Song2019WheresMD}}. 
The second strategy involves imputing the missing data values~\cite{Miao2023AnES, Sarma2022EvaluatingTU}. Methods such as constant, mean, or linear interpolation are used for imputation, followed by visualizing the uncertainty of these values. This is achieved through uncertainty visualizations~\cite{Hullman2019InPO, Kamal2021RecentAA, Kale2019HypotheticalOP, Correll2014ErrorBC, Kay2016WhenI, MacEachren2012VisualS}, including error bars, probability density plots (PDF), and confidence intervals (CI). 

Our technique combines two approaches - direct representation and uncertainty visualization, effectively displaying missing data in our inferred data tables. Specifically, our method highlights a visualization area where there are missing data points. Additionally, we visualize the data points whose values are inferred from the text. Thus, these inferred data points are inherently not certain. This dual strategy ensures a transparent and comprehensive analysis, maintaining the integrity and depth of the data interpretation.

%% file: src/3-method.tex
\section{Method}
\begin{figure*}[htbp]
    \centering
    \includegraphics[width=\linewidth]{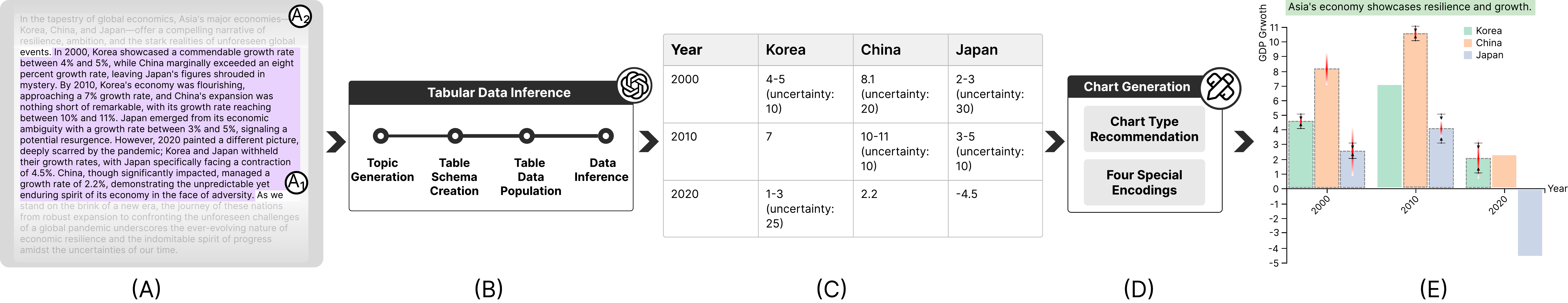}
    \caption{\textbf{The overview of transforming text into a chart.} The process begins with (A) Inputs which contains: A\textsubscript{1} the text statement users select and A\textsubscript{2} the context of the text statement. (B) The Tabular Data Inference transforms the text statement into a data table in 4 steps. (C) The resulting data table. (D) The appropriate chart type is recommended based on the characteristics of the tabular, and the chart is augmented with special visual encodings to accurately present the underlying data. (E) A generated chart represents the selected text statement.}
    \label{fig:method_overview}
\end{figure*}

\subsection{Overview}
Our method transforms data-involved text into 
\alexin{intuitive} charts, bridging the gap between textual data and 
\alexin{charts}. 
\alexin{The tool takes in two inputs: a text statement, which is a specific sentence of interest (Fig.~\ref{fig:method_overview} A\textsubscript{1}), and a context, which is the full text document (Fig.~\ref{fig:method_overview} A\textsubscript{2}).}
The output is the chart (Fig.~\ref{fig:method_overview} E) which visually presents the information in the text statement.
However, the transformation processing is not straightforward due to the complexity of data-involved texts, which contain \textit{text ambiguity}, \textit{data sparsity}, and \textit{subjective sentiment}. 
The complexities of text data pose specific challenges in transformation, including \textbf{uncertainties} in 
\alexin{text}, \textbf{data ranges}, \textbf{missing data values}, and \textbf{subjective sentiment} of the author.

Our method tackles these challenges through two key modules: tabular data inference and expressive chart generation. As delineated in Fig.~\ref{fig:method_overview}, when users read a data-involved text, they may select a 
statement (\alexin{Fig.~\ref{fig:method_overview}} A\textsubscript{1}) that they are interested in. 
\alexin{Then, the selected statement and context will be fed into \tool{}. \tool{} will analyze them, extract relevant information from them, and organize this information in a tabular format (Fig.~\ref{fig:method_overview} B and C).}
Based on the characteristics of the tabular data, \tool{} recommends chart types and designs special visual encodings to augment the chart (\alexin{Fig.~\ref{fig:method_overview}} D). The resulting chart (\alexin{Fig.~\ref{fig:method_overview}} E) accurately conveys the 
\alexin{underlying information of the statement.}
The LLM used in \tool{} is GPT-4 due to its superior reasoning ability~\cite{Achiam2023GPT4TR}.



\subsection{Tabular Data Inference}\label{method:section}
The primary objective of 
\alexin{tabular data inference} is to analyze a text statement 
, quantify textual information relevant to the statement, and transform it into tabular data.
We convert text into table format due to its clear structure and ability to handle diverse data types. This is ideal for our case, where text content exhibits various characteristics. 
\alexin{Tabular Data Inference encompasses four steps: topic generation (Section~\ref{method:topic_genertion}), table schema creation (Section~\ref{method:table_scheam_creation}, table data population, (Section~\ref{method:table_data_population}), and missing data estimation (Section~\ref{method:missing_data_estimation}).}
\alexin{\alexin{These steps result in a detailed table (Fig.~\ref{fig:method_overview} (C))}} that quantifies various textual characteristics, paving the way for subsequent chart generation.

\begin{figure}[htbp]
    \centering
    \includegraphics[width=\linewidth]{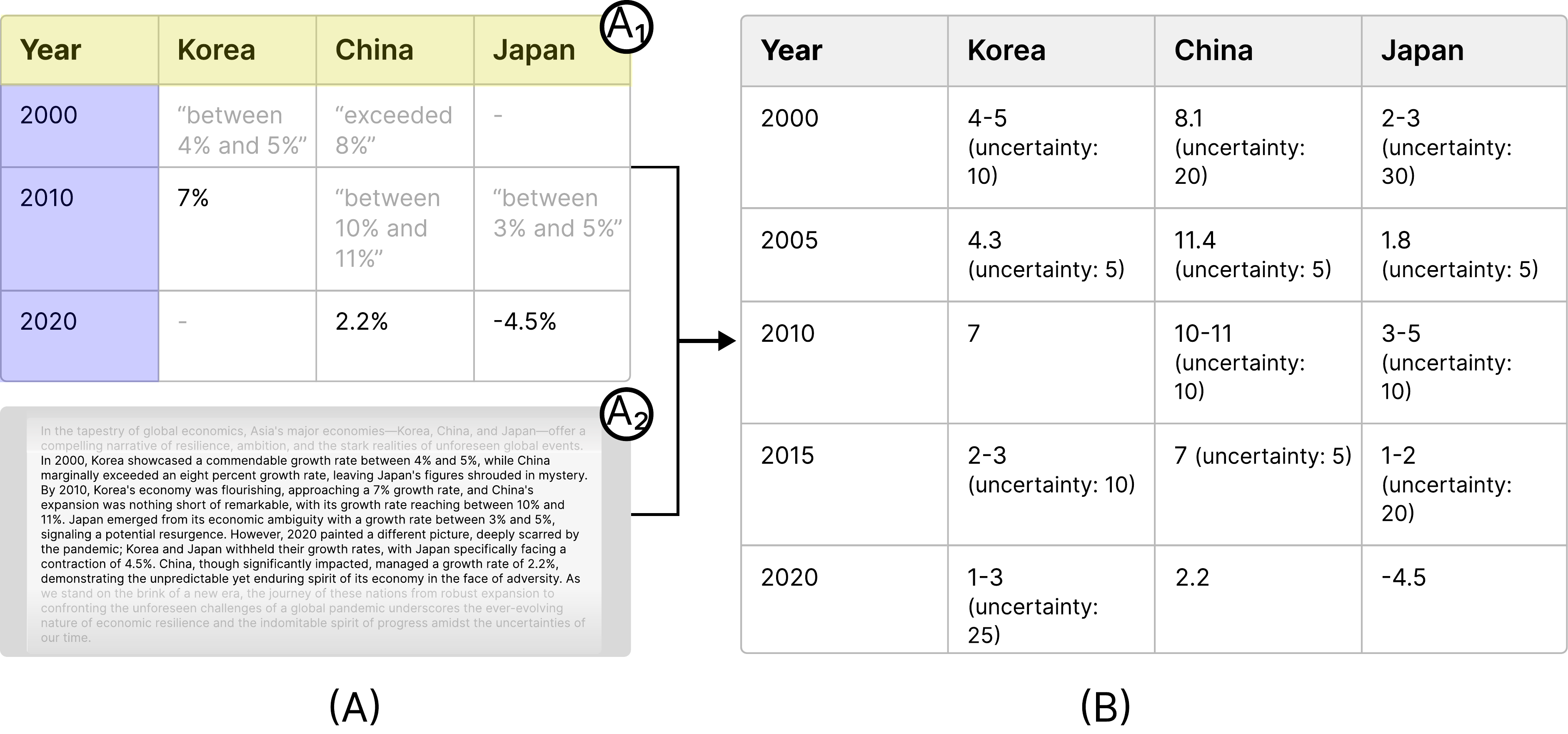}
    \caption{\textbf{Value Inference in Data Table:} (A) In A\textsubscript{1}, yellow highlights indicate the header row, while purple highlights mark row identifiers, both generated during Table Schema Creation. Cells with downplayed values indicate non-convertible data; empty cells signify missing values not directly found in the text statement. A\textsubscript{2} denotes the inputs: text statement and its context. (B) After the Data Inference, the table is completed, with inferred values assigned uncertainty scores to reflect confidence levels.}
    \label{fig:method_inference}
\end{figure}

\subsubsection{Topics Generation}\label{method:topic_genertion}
\alexin{Topic generation extracts topics from the text statement.}
\alexin{These topics help in selecting the appropriate data from the text. They direct \tool{} to select data accurately representing the intended message in the text statement.}
\alexin{To this end, \tool{} initially identifies and categorizes the crucial information within the statement by extracting key messages from the statement. These key messages provide a fundamental understanding of the textual content.}
As in the example text statement (Fig.~\ref{fig:method_overview} A\textsubscript{1}), we can obtain some key messages, e.g., 
\alexin{in 2000, Korea's growth rate was between 4\% and 5\%, China growth rate was larger than 8\%, and Japan's growth rate was between 3-5\% in 2010.}
To organize the extracted messages,
\alexin{\tool{} groups them} into different clusters \alexin{where each cluster presents a unique topic in the statement}. 
\alexin{To further ensure topics are both comprehensive and focused, \tool{} applies dual approaches, i.e., two clustering strategies to key messages. The two strategies are instructed to cluster key messages into a fine-grained level and a coarse-grained level. respectively. For example, given a statement (as depicted in Fig.~\ref{fig:method_overview} A\textsubscript{1}, fine-grained clustering can result in detailed topics, e.g., China's GDP growth rates changes in the 20 years, while coarse-grained clustering can provide a broader overview, e.g., Asian countries' general economic development trend.}

\subsubsection{Table Schema Creation}\label{method:table_scheam_creation}
\alexin{After obtaining topics, \tool{} organizes unstructured textual information into a structured format by developing a tailored table schema for each topic.}
\alexin{Each schema represents a topic and provides a dedicated template. It comprises a header row and row identifiers.}
The header row (depicted in yellow highlighted areas in Fig.~\ref{fig:method_inference} A\textsubscript{1}), labels each column with titles that describe the columnar data. Row identifiers (depicted in purple highlighted areas in Fig.~\ref{fig:method_inference} A\textsubscript{1}.) label each data row.  
\alexin{Header rows and row identifiers provide clear labels for the data in a table, indicating what types of data are required in the table.}
These labels help to guide the population of data and ensure the accuracy and integrity of the data are maintained~\cite{Wu2021TexttoTableAN, Li2023ASM, Pietruszka2022STableTG}. This is crucial for the next steps in data processing.

\alexin{In practice, we designed a prompt to instruct \tool{} to analyze a topic (i.e., a cluster of key messages obtained in Sectioned~\ref{method:topic_genertion}), and generate a schema for the topic. Specifically, \tool{} will identify the crucial phrases from the key messages. These phrases represent the core concepts of the topic.}
\alexin{To be specific, they} usually consist of nouns or noun phrases that convey the necessary information about the 
\alexin{key messages  }, such as "GDP Growth" and "Country" in an economic report.
Once the key phrases are identified, they are assigned to the header rows and row identifiers of a table schema. 
For example, if a key message discusses a company's "annual revenue", \tool{} will create a header row named 'Annual Revenue'.
Additionally, we have employed in-context learning~\cite{Dong2022ASO} to refine the process further by providing examples of key messages and their well-structured table schemas. This process allows \tool{} to learn from the examples, and thus generate a new schema that closely aligns with the key messages~\cite{Wang2023LLM4VisEV}. 

\subsubsection{Table Data Population}\label{method:table_data_population} 
\alexin{With a table schema established, \tool{} performs data population by extracting data from the context to fill the schema.}
\alexin{The data population fills the schema with quotes from the input context and then converts these quotes into numerical values. Quotes are phrases that contain explicit data and are directly taken from the context.}
\alexin{For example, a phrase ``7\%'' (depicted in Fig.~\ref{fig:method_overview} A\textsubscript{1}) denotes Korea's GDP growth in 2010, which corresponds to specific row identifier (i.e., 2010) and header row (i.e., Korea). Therefore, the phrase is quoted in the schema, as shown in the Fig.~\ref{fig:method_inference} A\textsubscript{1}.}
\alexin{Subsequently, the quote is converted to a numerical value ``7'' by \tool{}, as shown in Fig.~\ref{fig:method_inference} B. The reason the step includes the two-stage approach is that the approach can mitigate the LLMs' tendency to generate erroneous or fabricated information, known as the hallucination problem~\cite{Ji2023TowardsML}. However, requiring LLMs to reference specific information from the text for their answers can significantly reduce such inaccuracies~\cite{refrence_text}. \tool{} firstly quotes the original phrase from the context and then converts quotes to numerical values, since the desired table from \tool{} can accurately reflect the information in the text. To ensure the data population strictly follows our instructions, we used in-context learning. We have some pairs of inputs (i.e., table schemas and contexts) and valid output (populated tables) as examples in the prompt. These examples enable \tool{} to accurately interpret and replicate the desired data structuring process in new inputs.}

\alexin{Although \tool{} can find out data from the context, the resulting table is not perfect. As shown in Fig.~\ref{fig:method_inference} A\textsubscript{1}, the populated table contains quotes that cannot be directly converted to numerical values (e.g., ``exceeded 8\% '') and empty cells. These problems come from text ambiguity and data sparsity. Thus \tool{} hardly obtains the values of ambiguous quotes or even cannot find the data relevant to the schema.}

\subsubsection{Data Inference}\label{method:missing_data_estimation}

\alexin{Despite challenges in table data population, \tool{} aeffectively leverages the reasoning capabilities of LLMs, which include commonsense reasoning, arithmetic reasoning, and natural language understanding~\cite{Yang2023HarnessingTP, Huang2022TowardsRI,Achiam2023GPT4TR}.  These abilities enable [clarify what abilities reach the example result]. enable \tool{} to to precisely interpret ambiguous text and infer logical data values  For instance, it can deduce that a phrase like ``exceeded 8\% growth rate'' specifically refers to ``8.1\%'', or interpret ``between 4\% and 5\%'' to a range of ``[4-5]\%''.}

\alexin{\tool{}'s comprehensive understanding of textual content also allows it to address missing data points in tables. For example, if Korea’s 2020 GDP growth figure is absent (as shown in Fig.~\ref{fig:method_inference} A\textsubscript{1}). \tool{} identifies relevant contextual information to logically infer this missing value. Moreover, \tool{}  augmenting tables with additional data: as illustrated in Fig.~\ref{fig:method_inference} A and B, while the original table (Fig.~\ref{fig:method_inference} A\textsubscript{1}) displays economic growth for three countries over three years, the enhanced table includes data for five years (Fig.~\ref{fig:method_inference} B). This extension incorporates GDP figures for 2005 and 2015, discovered through \tool{}’s analysis of the text discussing these years. The new rows also maintain the structure of the original table.}


\alexin{Given the inherent ambiguity in the text, \tool{} evaluates the accuracy of each inferred result by assigning an uncertainty score ranging from 0 to 100. This score indicates the level of confidence in how well the inferred data matches the text's intended meaning. For instance, an explicitly stated data in the text, such as a specific figure, would receive an uncertainty score of 0, reflecting complete confidence. In contrast, inferred data based on less clear text, such as descriptions of a data range, might receive a score like 10, indicating moderate confidence. This scoring system helps users gauge the reliability of the information presented in the charts generated by \tool{}.}


\subsection{Expressive Chart Generation}

\alexin{After obtaining an inferred data table (Fig.~\ref{fig:method_overview} C) from text, \tool{} will visualize the table in a chart. \tool{} automatically selects an appropriate chart type based on the data's characteristics. Additionally, due nature of the text, the extraction of data encounters challenges from the text: \textbf{data ranges}, \textbf{uncertainties}, and \textbf{missing data values}. \tool{} will integrate these aspects into the chart to enhance data interpretation. Furthermore, \tool{} identifies the \textbf{subjective sentiment} in the text and also incorporates it into the chart. In sum, \tool{} generates a chart that enables users to glean a complete understanding of the text content directly from the chart.}


\subsubsection{Chart Type Recommendation}

\alexin{\tool{} will select axes and a chart type for the table because LLMs are effective in chart recommendation~\cite{Wang2023LLM4VisEV}. LLMs own the knowledge of visualization design. For example, LLMs recommend a line chart for the time-series data and a bar chart for categorical data. These recommendations adhere to visualization design guidance~\cite{munzner2014visualization}. Inspired by Wang~\textit{et al.} work~\cite{Wang2023LLM4VisEV}, \tool{} specifically selects a chart type for the table from four chart types: bar chart, line chart, pie chart, and scatter plot. These chart types are commonly used in visualization practices and suitable for a wide variety of data~\cite{Battle2017BeagleAE}. In practice, \tool{} recommends a chart type and axes in 3 steps. Firstly, \tool{} analyzes the table and recognizes its characteristics (e.g., data type). Secondly, it selects a chart type that is suitable for the table characteristics. Thirdly, \tool{} reviews the table to determine specific rows and columns as the axes of the chart type. For example, it will place time-related data on the x-axis for a line chart or categorical data on the x-axis for a bar chart. In sum, \tool{} specifies a chart type and axes for the table.}


\subsubsection{Special Visual Encodings}

\alexin{After determining the appropriate chart type and axes for the data table, \tool{} enhances the visualization with four specialized encodings. These encodings address the main challenges in textual data extraction: uncertainties, data ranges, missing values, and subjective sentiment. Each encoding is specifically designed to tackle one of these issues, ensuring that the visual representation is both comprehensive and precise. This approach allows \tool{} to convey the textual information effectively, with each encoding directly corresponding to a particular challenge:}

 
\begin{figure}[htbp]
    \centering
    \includegraphics[width=\linewidth]{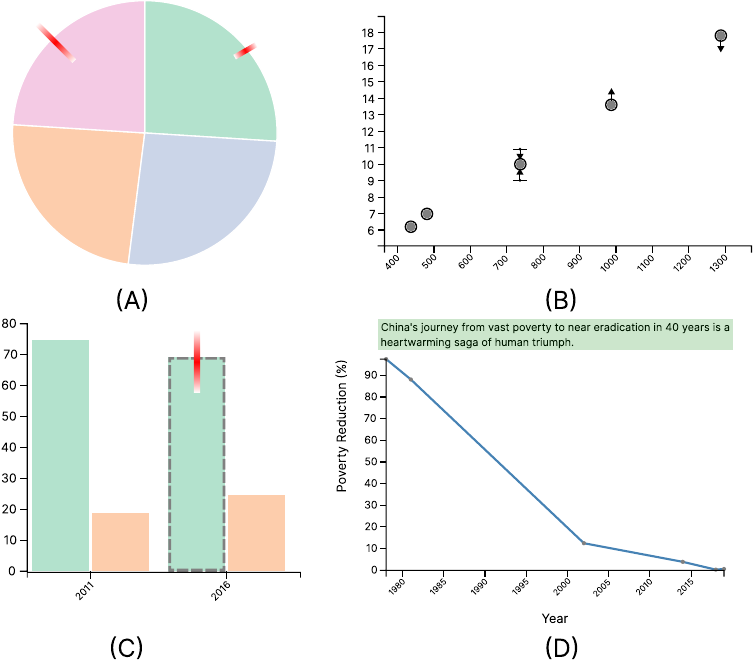}
    \caption{\textbf{Special Encoding Designs.} (A) Uncertainty Encoding represents the degree of uncertainty of data value. Longer stripe means larger uncertainty (B) Data Range Encoding represents data ranges in two conditions. Data value is within a specific range where its maximum and minimum values are determined; data value is either smaller or larger than a specific Fig.. (C) represents values that may not be in the text. (D) Sentiment encoding uses text annotation to describe the topic and uses background colors to represent positive, negative, and neutral sentiments, respectively.}
    \label{fig:encoding_method}
\end{figure}


\alexin{\textbf{Uncertainty Encoding} clarifies the uncertainty level in the data points. Due to the text ambiguity, a data point inferred from the text is not certainly precise. \tool{} assigns an uncertainty score to the data point in the tabular data inference (Section~\ref{method:section}). To visually signify the uncertainty score, \tool{} utilize the uncertainty encoding through gradient stripes (\raisebox{-0.2\height}{\includegraphics[scale=0.9]{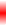}}). As shown in Fig.~\ref{fig:encoding_method}(A), a gradient is attached to a visual mark whose values are inferred from the text. Furthermore, the length of a stripe directly corresponds to the data point's uncertainty value: a longer gradient stripe indicates higher uncertainty. For example, if a visual mark with a higher stripe than other visual marks, this phenomenon implies the mark' values are less uncertain than other marks. This visual method provides an intuitive way to assess the reliability of each data point and the overall ambiguity present within the text, thereby enhancing the user's understanding of the data’s veracity.}


\alexin{\textbf{Data Range Encoding} visually depicts a value range of a data point on a chart.} \
\alexin{This technique encodes a} data that varies within a range rather than 
\alexin{a specific value.} 
\alexin{The data point without a specific value can be categorized} into two 
\alexin{types}: 
\alexin{closed-ended} ranges, such as `[9,11]', and open-ended ranges, like `>5' or `<5'. Consequently, 
\alexin{\tool{} uses} two distinct encoding strategies to accurately depict 
\alexin{them.} 
As shown in Fig.~\ref{fig:encoding_method} (B), 
\alexin{given a data value varies within a specific range [9,11]}, 
\alexin{\tool{}} calculate\alexin{s} the mean value from 
\alexin{its} two ends of the range (e.g., 9 and 11 in [9,11]), and placed the calculated value (e.g., 10) on the y-axis, with caps at both ends. Additionally, 
\alexin{\tool{} utilizes} two arrows towards the average value to indicate the indeterminate data range (\raisebox{-0.2\height}{\includegraphics[scale=0.6]{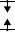}}).
Alternatively, also shown in Fig.~\ref{fig:encoding_method} (B), for open-ended ranges, 
\alexin{\tool{} uses} an arrow (\raisebox{-0.1\height}{\includegraphics[scale=0.9]{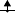}}) to indicate whether values surpass or fall below a specific value, clearly communicating the direction and extent of data variability.


\alexin{\textbf{Missing Value Encoding} identifies a data point whose value is not explicitly mentioned in the text. If the text mentions the data point but does not explicitly state its value, \tool{} will infer its missing value based on the contextual information. To visually indicate which values are inferred rather directly supported by the text, \tool{} employs a distinct missing value encoding, i.e., a dashed line is placed around the visual mark of the missing value data point (\raisebox{-0.2\height}{\includegraphics[scale=0.5]{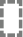}})~\cite{Boukhelifa2012EvaluatingSA} on the chart. This encoding serves as a visual cue to users, highlighting that the data is derived from interpretation rather than direct extraction. Therefore, missing value encoding enhances transparency and aiding in data interpretation reliability.}

\alexin{\textbf{Sentiment Encoding} captures the emotional tone of the text and associates the emotion with data points. \tool{} analyzes the text statement, identifies the author's subjective sentiment from the text, and visualizes it on the chart. \tool{} finish the task in 4 steps. First of all, \tool{} analyzes a topic (obtained in Section~\ref{method:topic_genertion}), and classifies its sentiment as negative, neutral, or positive. Secondly, following the classification, \tool{} identifies data points related to this sentiment. Thirdly, \tool{} generates a narrative to describe the set of data points. The narrative is concise and emotional; it provides an external description to link the data points and the sentiment. Fourthly, The narrative is visualized on the chart as a text annotation. Text annotation can support users' understanding of the chart~\cite{Lundgard2021AccessibleVV}. Guided by Stokes~\textit{et al.}'s work~\cite{Stokes2022StrikingAB}, \tool{} specifies a rule of the annotation placement: the annotation describes a single data points will be placed near the single data point (shown in Fig.~\ref{fig:teaser} A\textsubscript{2}); otherwise, it will be placed in the title to describe the chart's overall pattern  (shown in Fig.~\ref{fig:encoding_method} D). The rule enables users to better link the data point with the text or perceive the chart's pattern. Furthermore, \tool{} incorporates the color in the sentiment narrative because color can enable users to better perceive the sentiment from the chart~\cite{Bartram2017AffectiveCI}. Specifically, inspired by Batram \textit{et al.}'s work~\cite{Bartram2017AffectiveCI}, \tool{} use different color backgrounds in the sentiment narrative, i.e., positive sentiment (\raisebox{-0.3\height}{\includegraphics[scale=0.3]{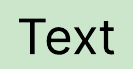}}), negative sentiment (\raisebox{-0.3\height}{\includegraphics[scale=0.3]{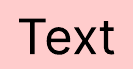}}) and neutral sentiment  (\raisebox{-0.3\height}{\includegraphics[scale=0.3]{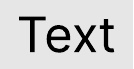}}). As a result, the sentiment encoding can express the sentiment of the text and provide an external description to the chart.}

%% file: src/4-evaluation.tex
\section{Case Study}
In our case study, \tool{} was applied to generate bar charts, line graphs, pie charts, and scatter plots from three real-world documents—a scientific paper, a survey report, and a news article—covering health, religion, and politics. Fig.~\ref{fig:teaser} displays these \alexin{documents'} excerpts with the resulting charts. 
\alexin{These examples illustrate how \tool{} enhances users' reading experience by transforming texts into informative charts.}

\textbf{A Scientific Paper about Covid-19}\cite{Rosenstrom2020HighQualityMR}
discusses the effects of high-quality masks adoption in reducing Covid-19 infections and deaths in the US. When a user examines the paper is immediately drawn to its assertive title (red text in Fig.~\ref{fig:teaser} A\textsubscript{1}). The user visualizes this title through \tool{}, which interprets the title and extracts relevant data from the document. Despite not reading the entire paper, the user can see the data related to the title represented in a bar chart (Fig.~\ref{fig:teaser} A\textsubscript{2}). The data are derived from pertinent excerpts in the document. For example, one excerpt mentions different mask adoption rates and their effects in preventing COVID-19 in the last paragraph of the document (black text in Fig.~\ref{fig:teaser} A\textsubscript{1}), and these data are presented in the bar chart. As 
\alexin{the user} examine the chart, they first notice the text annotation reflecting the subjective sentiment about mask adoption and its impact on infections, deaths, and prevalence rates. The bars on the chart delineate COVID-19 reduction rates across these metrics, clearly showing that higher mask adoption will lead to lower infection impact. Data range encoding (\raisebox{-0.1\height}{\includegraphics[scale=0.9]{figs/open_data_range.pdf}}) suggests that the benefit of 50\% mask adoption could be greater than depicted,  while uncertainty encodings (\raisebox{-0.2\height}{\includegraphics[scale=0.9]{figs/graident_stripe.pdf}}) might be less reliable due to lack of explicit numerical information in the document. 

Further reading leads the user to a data-involved paragraph (blue text in Fig.~\ref{fig:teaser} A\textsubscript{1}), which \tool{} transforms into a scatter plot (Fig.~\ref{fig:teaser} A\textsubscript{3}). Through sentiment encoding~(\raisebox{-0.3\height}{\includegraphics[scale=0.3]{figs/positve_sentiment.pdf}}), the user perceives the strong effects of mask adoption. The scatter plot clearly ranks the different strategies dealing with COVID-19, confirming that masks significantly reduce death and hospitalization. Mobility changes alone are less effective, and the strategy, without masks and no interventions, is the least effective. Encoding for missing values (\raisebox{-0.2\height}{\includegraphics[scale=0.5]{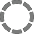}}) and uncertainty aid the user in discerning data missing in the text and the reliability of inferred results.


\textbf{A Politic News Article}\cite{trump}
examines President Trump's first 100 days, a user begins by reading a descriptive paragraph (
red text in Fig.~\ref{fig:teaser} B\textsubscript{1}), which states without specific data. Intrigued by the statement regarding Trump's popularity, the user turns to \tool{} for deeper analysis. \tool{} scrutinizes the document, identifying relevant data on Trump’s support rates from the entire article and representing this information visually in a bar chart (Fig.~\ref{fig:teaser} B\textsubscript{2}). This chart delineates varying public attitudes toward the administration, and the sentiment encoding~(\raisebox{-0.3\height}{\includegraphics[scale=0.3]{figs/negative_sentiment.pdf}}) reveals the negative sentiment within the article: the public generally disapproves of Trump's administration. Shifting to more specific information, the user obtains a comprehensive view of public opinion from the bar chart. Encodings for uncertainty (\raisebox{-0.2\height}{\includegraphics[scale=0.9]{figs/graident_stripe.pdf}}) and missing data (\raisebox{-0.2\height}{\includegraphics[scale=0.5]{figs/dashed_bar.pdf}}) signal that some data in the chart are inferred values rather exact numbers in the article. For example, data in ``Disregards Public Worries'' (the rightmost bar in Fig.~\ref{fig:teaser} B\textsubscript{2}) is described as \textit{``nearly 6 in 10''} in the article.

The user continues the reading and encounters a data-rich statement (
\alexin{blue text in} Fig.~\ref{fig:teaser} B\textsubscript{1}). The statement compares Trump's approval ratings with past presidents. \tool{} can generate a chart that uses a bar chart to visualize Trump's ratings compared with those of Bill Clinton and Barack Obama. Missing value encoding indicates an inferred disapproval rating for Clinton. The inferred rating is derived from the comparison to Trump's disapproval rating, i.e., \textit{``Trump's disapproval rating is 14 percent higher than Bill Clinton's''}. Therefore, \tool{} performs an algebraic operation to infer Clinton's disapproval rating. Through \tool{}, the user can navigate complex political news from informative charts.

\textbf{A Surveyed Report about U.S. Religious Demographic }\cite{religiously_unaffiliated} delineates the shifting religious composition in the United States.  A user navigates the report and is first presented with an introductory statement (
\alexin{red text} in Fig.~\ref{fig:teaser} C\textsubscript{1}). The sentence lays the groundwork for understanding the nation’s evolving belief trend. This initial statement, while rich in qualitative insights into the secularization of U.S. society, lacks numerical data. To visualize these abstract descriptions, the user turns to \tool{}. \tool{} interprets this qualitative sentence and, as illustrated in Fig.~\ref{fig:teaser} C\textsubscript{2}, utilizes sentiment encoding to translate the author's neutral tone (\raisebox{-0.3\height}{\includegraphics[scale=0.3]{figs/neutral_sentiment.pdf}}), which refers to the factual decrease of self-identified Christians to 63\% and a rise in religious `nones' to 29\%. These 
\alexin{statements} are provided by the report's content, stating a 29\% composition of religious `nones' among U.S. adults, while Christians constitute 63\%, as seen in black text in Fig.~\ref{fig:teaser} C\textsubscript{1}. The user then explores the trends over time depicted in a line chart (Fig.~\ref{fig:teaser} C\textsubscript{2}). At a glance,  the chart delineates a discernible pattern: a consistent decrease in religious affiliation contrasted against a rise in secularism. Importantly, the user can notice that the line chart's data points do not contain any missing value or uncertainty encoding, which means these data values are unequivocally articulated in the report. As a result, the user is convinced of the credibility and validity of the information presented in the chart.

Seeking a more granular perspective, the user delves into the report's section with a data-involved statement (
\alexin{blue text} in Fig.~\ref{fig:teaser} C\textsubscript{1}). The statement offers specific statistics about the demographic of White Protestants. \tool{} translates this information into a pie chart, segmenting the data to reflect the different religious affiliations. Similar to the line chart, there are no encodings of missing values and uncertainty, and thus the user knows that the data value in the chart explicitly exists in the report, thereby reinforcing 
\alexin{the user's} confidence in the information conveyed in the pie chart. Consequently, \tool{} enables the user to gain an immediate and comprehensive understanding of the multifaceted religious composition within the U.S.

\section{Expert Interview}
We conducted expert interviews to evaluate the quality of charts generated by \tool{}. To this end, we engaged three experts with experience in data visualization to obtain their feedback and insights into the functionality of \tool{}. The experts (E1-E3) have been involved in data visualization research for at least one year. E1 was a Ph.D. student. E2-E3 have master's degrees. All experts have published at least one paper in the top journals of data visualization. The interviews were conducted via Zoom due to geographical limitations, with each session lasting approximately one hour.

\alexin{Expert interview evaluates the quality of charts generated by \tool{}. \tool{} should ensure these charts can accurately and intuitively represent the textual information. To this end, we invited three experts in data visualization to evaluate whether the charts can meet the requirement and also collected their insights about \tool{}. The experts (E1-E3) have at least 1 year of experience in data visualization. E1 is a Ph.D. Candidate and E2-E3 have master's degrees. All experts have published at least one paper in the top visualization journals. We experimented with the experts via Zoom because we and the experts were in different regions. We experimented with experts respectively. Each experiment session lasted for 1 hour.}

\subsection{Tasks}
\alexin{Experts evaluated charts' quality from four aspects.}
The experts were presented with charts alongside their corresponding text statements and contexts (i.e., excerpts from documents). Text statements and contexts are inputs of \tool{}, and their outputs are charts used in the task. 
\alexin{Experts reviewed these charts, text statements, and context. Then they assigned scores in four key metrics:}

\begin{enumerate}

    \item Relevance between Text and Chart:  
    \alexin{it assesses whether the \tool{} can generate a chart whose information is relevant to a selected text. The scores ranges from 1 (least relevant) to 5 (most relevant).}
    \item  Data Accuracy
    \alexin{: it evaluates whether \tool{} can accurately extract the data from the text and visualize them on charts. The metric scores from 1 (least accurate) to 5 (most accurate).}
    \item  Chart Clarity
    \alexin{: it tests whether experts can clearly understand the message conveyed by the charts. The metric scores from 1 (least clear) to 5 (most clear).}
    \item  Visualization Guideline Compliance
    \alexin{: it examines whether \tool{} can recommend chart types compliant with the guidelines in data visualization. The metric scores from 1 (least compliant) to 5 (most compliant).}

\end{enumerate}

\subsection{Material \& Procedure}
In our expert interview, we presented 10 documents that were also used in the user study, along with 20 charts. Each document had 
\alexin{two} charts, each with the same text statement and context. This was because a single text statement can have multiple topics that have their own chart. The charts were randomly selected. The charts were exported as images with resolutions 500 $\times$ 500 \alexin{(pixels)}.

At the beginning of the 
\alexin{experiment}, each expert was given a brief 10-minute introduction about 
\alexin{our method and the experiment instructions}. Then, they were asked to complete 
\alexin{tasks} in 
online 
\alexin{questionnaires} using Qualtrics\footnote{\url{https://www.qualtrics.com}}. After completing 
\alexin{tasks}, 
\alexin{we} conducted a semi-structured interview, during which 
\alexin{experts} 
were encouraged to provide feedback on the \tool{}'s strengths and weaknesses, as well as any features they found particularly helpful or unhelpful.


\begin{table}[]
\centering
\begin{tabular}{lll}
\hline
 \textit{\alexin{Metrics}} &  \textit{Mean}&  \textit{SD} \\ \hline
 Relevance of Text to Chart&  4.88&  0.41\\ \hline
 Data Accuracy.&  4.82&   0.53\\ \hline
 Chart Clarity &  4.70&  0.69\\ \hline
 Vis. Guideline Compliance&  4.57&  0.80\\ \hline
\end{tabular}
\caption{\alexin{Quality evaluations result } for the charts generated by \tool{}.}
\label{table: expert_rating}
\end{table}

\subsection{Result \& Analysis}

\textbf{Experts Ratings' Consistency.} \alexin{Experts' scores were consistent in all tasks.} Experts' evaluations across four key metrics were assessed for consistency using Randolph’s multi-rater kappa statistic~\cite{Randolph2005FreeMarginalMK}. Specifically, the metric \textit{Relevance between Text and Chart } and \textit{Data Accuracy} achieved a kappa of 0.771 and 0.692, indicating a substantial inter-rater agreement~\cite{Randolph2005FreeMarginalMK}. The \textit{Chart Clarity} and \textit{Vis. Guideline Compliance} metrics showed moderate agreement, with kappa values of 0.56 and 0.41, respectively~\cite{Randolph2005FreeMarginalMK}. These kappa scores demonstrate a reliable level of consensus among the experts.

\textbf{Chart Quality Assessment.} 
\tool{} can deliver high-quality charts. Table~\ref{table: expert_rating} illustrates that the \textit{Relevance between Text and Chart} achieves an average score of 4.8, approaching the ideal score of 5, which signifies high relevance between text statement and charts. This indicates that the charts successfully convey the essential messages of the text. Moreover, with an average score of 4.82 in \textit{Data Accuracy}, \tool{} demonstrates its ability to \alexin{accurately} quantify textual information for visual representation. 
Additionally, the average scores above 4.5 are in both \textit{Chart Clarity} and \textit{Vis. Guideline Compliance}. These metrics confirm that \tool{} not only creates charts that are easy to interpret but also recommends chart types that are suited to the data characteristics. Generally, these scores verify \tool{}'s performance in accurately extracting textual information and generating intuitive charts.

\subsection{Expert Feedback}
\alexin{We collected and analyzed experts' feedback. Then, we categorized their opinions in three parts:}

\textbf{Applications of \tool{}.} Experts collectively agree that \tool{} shows great promise in enhancing text comprehension, notably its proficiency in visualizing textual information in charts. They underscore the benefits of using charts for information acquisition, pointing out that charts significantly reduce cognitive load compared to navigating through dense, data-involved text. 
E3, mentioned, \textit{``\tool{} significantly eases my workload by aiding in the document reading process."} E3 also identified a key application of \tool{}: providing users with document overviews and guiding them to specific sections of interest. E1 further elaborated on this functionality, explaining,\textit{ ``For instance, when analyzing an essay, one could highlight a thesis statement. \tool{} then generates a chart visualizing the text content pertinent to the thesis. This approach offers a snapshot of the essay's supporting arguments.''} E1 also expressed optimism about \tool{}'s capabilities, categorizing its potential into two primary functions: 1.revealing 
\alexin{insights} within the text, and 2.extracting relevant data to bolster descriptive text, such as conclusions or arguments. \textit{``Users often have two objectives when analyzing textual data: exploring to uncover insights, like a company’s revenue trend over time, and verifying causation''} E1 noted. He clarified that \tool{} can effectively display 
\alexin{insights of a data-involved text.} Additionally, \tool{} can assist in causation verification which involves a user wanting to find evidence supporting a conclusive statement. E1 mentioned \textit{``Without the help of \tool{}, it would require me to meticulously review text. However, \tool{} simplifies this process by allowing me to select the conclusion, upon which the tool automatically gathers and analyzes the relevant context, converting it into a chart.''}  This innovative approach enables him to quickly 
\alexin{check the support evidence of the conclusion from} -charts, thereby saving time and reducing cognitive strain.

\textbf{Encoding Design.} Experts appreciated the \tool{} encodings for their ability to elucidate the characteristics of textual information. E1 particularly commended the inclusion of encodings for uncertainty and missing values. E1 clarified that while the encoded data points in charts are recognized as approximations rather than precise values, they significantly enhance the understanding of overall patterns within the text. Thus, superior to hidden data points or arbitrary data points, inferred data points offer substantial insight into the text. E1 provided a compelling example: \textit{``When examining a text that presents data on the racial makeup of a job market's workforce for a specific year, and the conclusion states that the racial composition has increased in the last decade, \tool{} generates a line chart with three ascending lines representing each racial group mentioned in the text. The chart for the last ten years on the line chart are inferred and encoded with missing or uncertain data. Although these charts may not be completely accurate, the line chart provides new perspectives that enhance the interpretation of the text.''} E3 also acknowledged the advantages of such encodings, with a particular interest in sentiment encodings. E3 noted that sentiment encoding not only presents key messages within the text but also links text with its chart. This connection enhances the comprehension of the text-chart relationship, making it easier to extract insights from the chart. E3 emphasized that sentiment encoding provides an additional layer of understanding, particularly in narratives such as news articles or essays that reflect the author's subject sentiment. It not only improves the comprehensibility of the chart but also makes it emotionally resonant.

\textbf{Challenges in \tool{}.} While \tool{} advances in data visualization, it is not without its limitations, particularly in the area of semantic grouping. A primary concern, as noted by experts E1 and E3, is the tool's approach to chart selection, which often overlooks the semantic meanings of the data being visualized.
E3 highlights a critical flaw with an illustrative example, noting, \textit{``I encountered a bar chart comparing the average longevity of a country's population with mortality numbers from various diseases. Despite both datasets being numerical and category-based, their juxtaposition in a single bar chart is conceptually inappropriate due to their fundamentally distinct semantic contexts.''} This instance underscores a key limitation in \tool{}'s chart recommendation algorithm, which, while adept at matching data types to appropriate chart forms, fails to account for the semantic significance of the data.
To address this issue, expert E2 proposes the integration of an additional module within \tool{}'s workflow, specifically designed for semantic grouping. This module would precede the chart recommendation step, categorizing data based on semantic meanings. E2 acknowledges that the success of this solution hinges on the model's ability to comprehend semantic nuances.
Offering an alternative strategy, E1 suggests leveraging the inherent semantic structure within documents. He explains, \textit{``Data related by semantic meaning are typically clustered together within a document, such as in adjacent paragraphs discussing related topics. By analyzing the document's structure and topic sentences, \tool{} could more accurately group data for visualization.''} This method leverages the document's existing organization to enhance the relevance and coherence of generated charts.
Experts E2 and E3 criticized the tool's simplistic sentiment representation through color encoding. E3 suggested incorporating
emoji icons
to convey a broader range of emotions more effectively.

\section{User Study}
For the user study, we aimed to ascertain whether \tool{} can assist users to swiftly derive insights from the text. 
%
%
%
our study will deliver a comprehensive analysis of how our method empowers users to extract and comprehend insights from text.

\subsection{Participants}
In the user study, we recruited 15 participants, with an age range from 26 to 28 years (mean=27, SD=1.83). The group comprised 5 females and 10 males, all of whom came from a local university and had at least 
\alexin{bachelor degree}. All of them are familiar with data visualization and use data visualization daily.
Prior to the participation, each participant signed an IRB-approved consent form and received \$15 for their contribution to our study.

\subsection{Apparatus \& Material}
\textbf{Apparatus. } To evaluate our method, participants engaged in the study remotely from their personal computers, simulating their usual document reading settings. The experiment was facilitated through Zoom, where participants completed a 
\alexin{questionnaire generated by Qualtrics}. They were required to share their screen during the survey for monitoring purposes. Throughout the experiment, participants had the opportunity to ask questions directly to the conductor using Zoom's microphone feature.
 
\textbf{Material.} 
This study utilized a set of 10 documents, representing a wide range of real-world documents across various fields including politics, economics, medicine, sports, and technology. These documents—comprising scientific reports, news articles, and academic papers—were chosen for their diversity and relevance. Given the lengthy nature of some documents, we opted to use excerpts for the evaluation. Each document provided a unique excerpt, resulting in 10 excerpts used to assess \tool{}'s ability to 
\alexin{help users better understand the textual information} across different types of documents. This approach ensures a thorough evaluation of \tool{}, highlighting its applicability across diverse text forms. 
To further refine our analysis, we selected 10 text statements from these excerpts, one per document. These statements, which range from containing explicit data to being purely descriptive, serve as the basis for chart generation by \tool{}, illustrating its versatility. The method inputs each document's text statement along with its excerpt as context, generating a chart that encapsulates the essence of the statement. This process not only tests \tool{}’s effectiveness in visual representation but also demonstrates how it contextualizes text within the broader narrative provided by the excerpts. In short, our user study incorporated 10 excerpts and 10 charts, each derived from one of the excerpts.



\subsection{Task \& Metrics}
\textbf{Task.} To comprehensively evaluate our proposed method, we conducted two tasks: assess \tool{} utility and validate encoding effectiveness. In \tool{} utility assessment, we assessed \tool{}’s effectiveness in enabling swift and accurate information extraction from texts. Participants were tasked with answering a set of multiple-choice questions— one question for each of the 10 documents provided. These questions, aligned with visualization tasks outlined in Tamra~\cite{Munzner2014VisualizationAA} (identify, compare, summarize), were constructed to gauge participants' comprehensive understanding of either the text or the accompanying chart. Successful answers required a thorough grasp of the document’s content, as participants chose responses based on data values identification, comparative analysis of values, or summarization of insight within the text or chart. The study was conducted under two conditions to evaluate \tool{}:

\begin{itemize}
    \item A baseline condition where participants answered questions using the text alone.
    \item A \tool{} condition where answers were derived from charts generated by our method.

\end{itemize}

The task design is within-subject design, which ensures each participant engaged with both text and chart formats, answering 5 questions under each condition. To mitigate potential memory and order effects, each question was seen only once by a participant, with a randomized order of questions. This approach guarantees a balanced evaluation of \tool{}'s utility in facilitating quick, accurate data interpretation, and compares its performance directly against traditional text analysis.

In addition, to validate the effectivness of our four visualization encodings on textual information comprehension, we conducted an ablation study. Our methodology involved comparing a baseline chart—lacking any of our specialized encodings—to four separate charts, each incorporating one unique encoding. Participants were provided with a text paragraph and asked to evaluate whether the chart with a specific encoding improved their understanding of the text-chart relationship over the baseline. This comparison was repeated for each encoding, allowing us to isolate and validate the effectiveness of our visualization strategies. Through this structured approach, we determined the contribution of each encoding to enhancing user comprehension.

\textbf{Metrics.} To rigorously evaluate the \tool{} framework, we employed four distinct metrics, each targeting a distinct and crucial aspect of text reading:
\begin{itemize}
    \item The correctness of the questions answers. The correctness of these answers reflects the method's capability to facilitate understanding. Notably, when a question allows multiple correct responses, we employ a scoring system based on the Jaccard index~\cite{real1996probabilistic}. This means the score for each question is calculated by dividing the number of correct answers selected by the participant by the total number of unique answers (both correct options and those chosen by the participant). For instance, if the correct answers are A, B, and C, and a participant selects A, C, and D, the score would be 2/4, representing the intersection (A and C) over the union (A, B, C, and D) of correct and selected answers.
    \item The time taken to answer each question. The time cost that each participant took to answer each question was recorded under the different experimental conditions. This metric is critical for assessing the efficiency of \tool{} framework, particularly in terms of reducing the time required for participants to extract and process information from the texts. 
    \item The cognitive workload experienced by participants. Participant's cognitive workload was measured using the NASA-TLX scale\alexin{~\cite{Hart1988DevelopmentON}}. This self-reported metric provided insight into the cognitive demands of each participant, offering a measure of the method's impact on user experience in terms of ease of understanding and cognitive load.
    \item The usefulness of our four designed encodings. We assessed the usefulness of our four encoding designs—
    \alexin{uncertainty encoding, data range encoding, missing value encoding, and sentiment encoding}—using a Likert scale from 1 (not useful) to 7 (
    useful). This scale helped evaluate participants' perceptions of how each encoding aids in accurately and confidently interpreting chart data and sentiments. 
    The goal was to quantitatively validate each encoding's role in enhancing users' comprehension and interpretability of text-chart relations.
\end{itemize}
In short, these metrics form a comprehensive set of criteria for evaluating the effectiveness of the \tool{} framework in enhancing the reading and comprehension of text.

\subsection{Procedures}
Each participant go through the following procedures in our user study.

\textbf{Introduction and Tutorial 
}  
\alexin{Participants} first filled out a demographic questionnaire and then received a tutorial explaining our method and the subsequent tasks.

\textbf{Utility Task
} 
Following the tutorial, participants engaged in answering multiple-choice questions without a time limit, based on either text excerpts or charts. This task was designed to test their ability to extract information accurately under two conditions—exclusively text or charts. Participants began with five questions in one condition before switching to the other, ensuring exposure to both scenarios. After each set, they assessed their experience using the NASA-TLX workload measure and provided feedback via a 
\alexin{7-point Likert scale (1-low, 7-high)}.

\textbf{Ablation Task
} Next, participants assessed the impact of our visualization encodings by comparing a baseline chart to ones with specific encodings. This evaluation aimed to determine the encodings’ effectiveness in enhancing text comprehension from chart. For each encoding, participants rated its usefulness on a 7-point scale, progressing through all encodings sequentially.

We also collected participants comments on \tool{} regarding its utility, design and clarity of the generated charts, which can be found in the supplementary material.


\subsection{Results \& Analysis}

\begin{figure}[htbp]
    \centering
    \includegraphics[width=\linewidth]{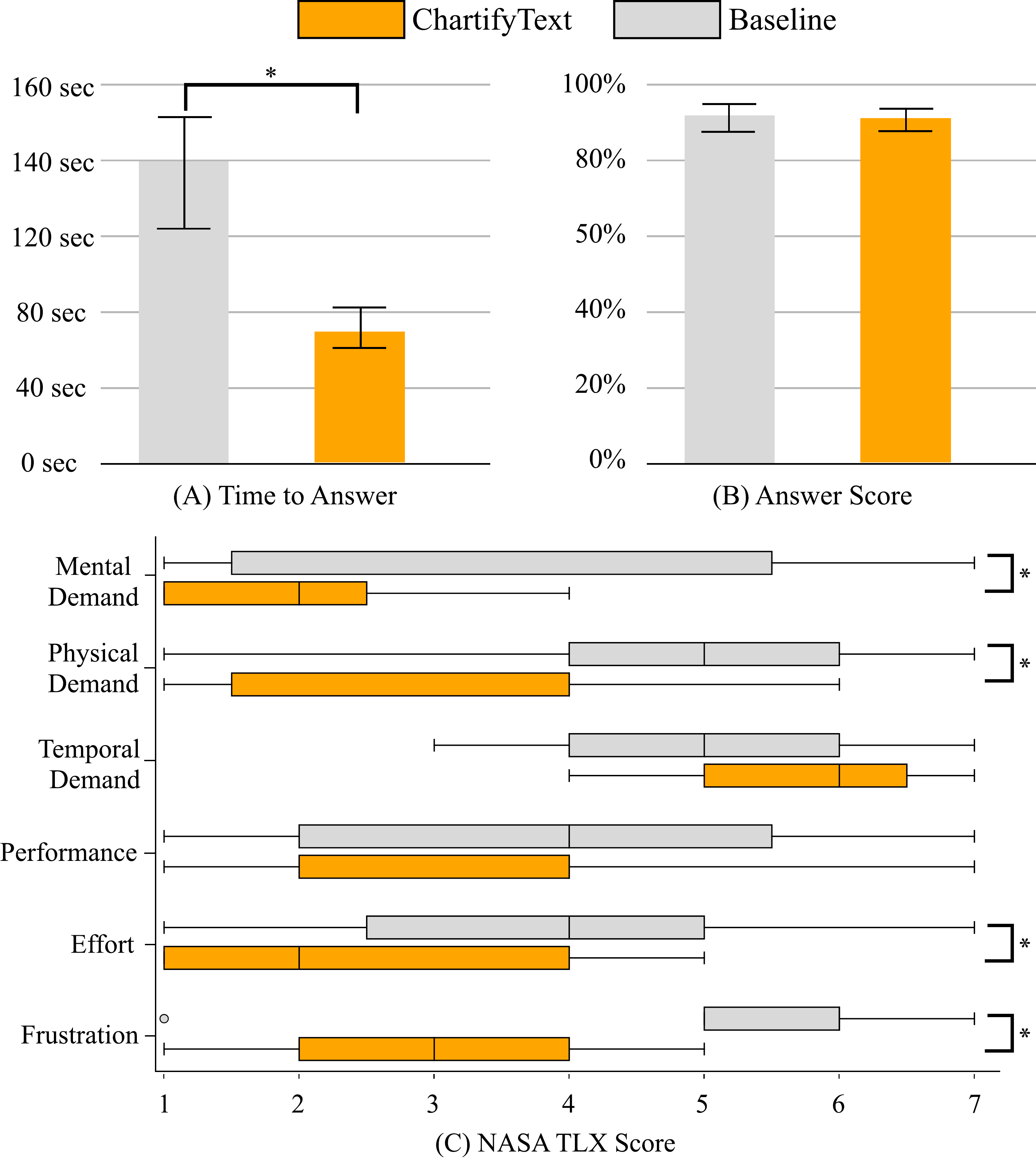}
    \caption{User study evaluation result. (A) The average time to answer a question. (B) The average score in the answering. (C) The NASA Task Load Index.}
    \label{fig:user_study_result}
\end{figure}


We first performed a normal distribution test on 
\alexin{result data}. Based on the distribution, we applied a repeated measures t-test for normally distributed data and a repeated Wilcoxon test for non-normal distributions.

\textbf{Time.} The result shows that using \tool{} resulted in significantly faster information acquisition than the baseline method. The repeated measures t-test reveals a significant difference between the average time of \tool{} and the baseline ($p$=0.00009) (Fig.~\ref{fig:user_study_result} A). Participants spent much less amount of time by \tool{} ($\mu$=73.62s, $\sigma$=9.46s) on answering than by the baseline ($\mu$=139.36s, $\sigma$=10.58s).

\textbf{Answer Score.} The result shows that using the \tool{} method resulted in the same level of accuracy in acquiring information as the baseline method. The repeated Wilcoxon test did not find a significant difference in the scores between the two conditions ($p$=0.7860) (Fig.~\ref{fig:user_study_result} B). Participants were able to achieve comparable scores when answering questions based on the information presented in \tool{} ($\mu$=90.00, $\sigma$=4.00) as they did when using the text ($\mu$=92.0, $\sigma$=5.00).

\textbf{Workload (NASA-TLX).} The result shows that \tool{} significantly reduces participants' workload in information acquisition with \tool{} than the baseline method (Fig.~\ref{fig:user_study_result} C). The repeated Wilcoxon test reveals a significant difference between \tool{} and baseline for mental demand ($p$<0.001), physical demand ($p$<0.007), and frustration ($p$<0.005), and does not find a significant effect between two conditions for performance  ($p$>0.6). The result of participants' self-reported performance is consistent with the result of their score in question answering. The repeated measures t-test reveals that there is a significant difference between the two conditions ($p$<0.009) \alexin{in effort} and not a significant difference in temporal demand ($p$>0.39).

\begin{figure}[htbp]
    \centering
    \includegraphics[width=\linewidth]{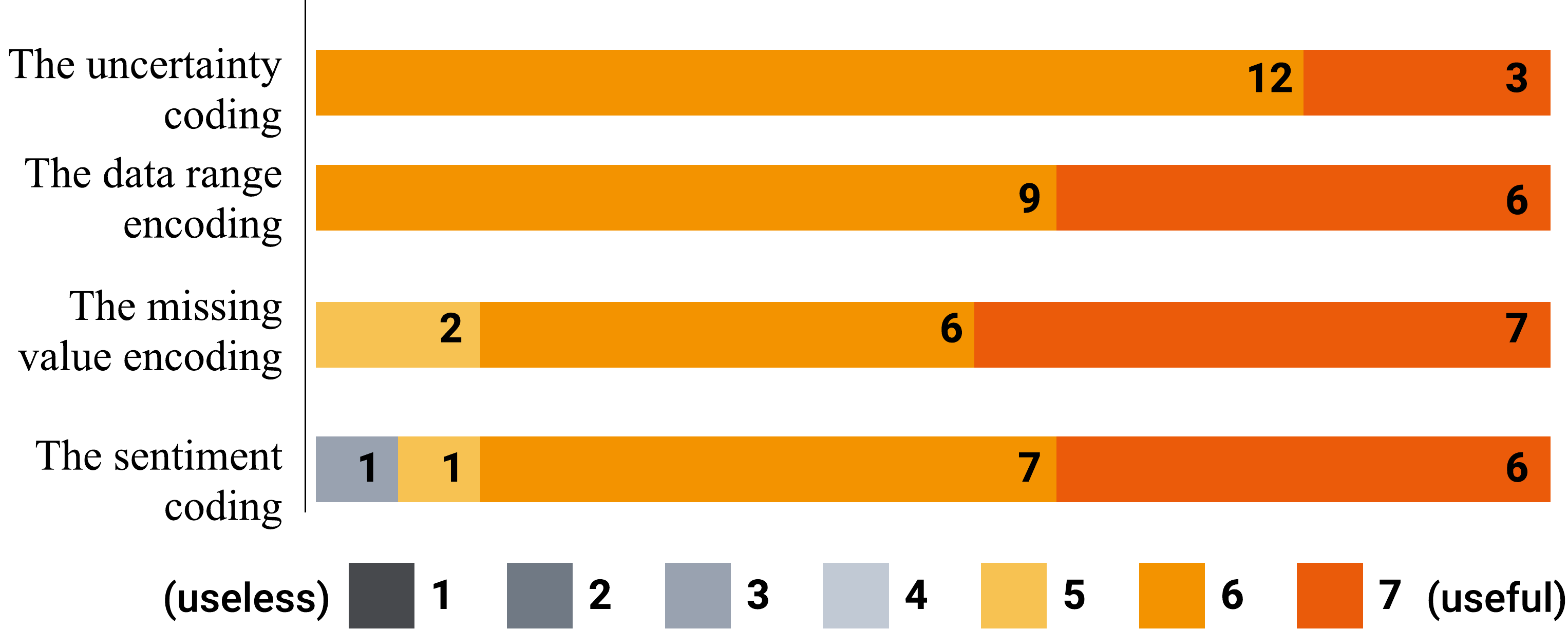}
    \caption{Participants' ratings for the four visual encodings on 7-point scale.}
    \label{fig:encoding_result}
\end{figure}

\textbf{Usefulness of encodings.} Most participants found the four encodings to be useful for enhancing their comprehension of textual information 
\alexin{,as shown} in Fig.~\ref{fig:encoding_result}
. In comparison to plain text without any of our designed encoding
, they found that the uncertainty encoding (\raisebox{-0.2\height}{\includegraphics[scale=0.9]{figs/graident_stripe.pdf}}) helped them understand the reliability of data points due to ambiguous text. The data range encoding (\raisebox{-0.1\height}{\includegraphics[scale=0.9]{figs/open_data_range.pdf}}) was also reported as useful by participants, which visually marked value spans on charts. The missing value encoding (\raisebox{-0.2\height}{\includegraphics[scale=0.5]{figs/dashed_circle.pdf}}) helped them identify the data points that were not included in the text statement. Finally, the sentiment encoding (\raisebox{-0.3\height}{\includegraphics[scale=0.3]{figs/negative_sentiment.pdf}}) helped most participants perceive the underlying emotions from the text statement and connect them with the data points in the chart.

%% file: src/5-discussion.tex
\section{Discussion}

\textbf{Limitation of GPT-4 Capability.} The \tool{} depends on the 
capability of GPT-4. 
\alexin{However,} GPT-4 has some limits. For example, they can only allow input text up to a certain length, which means they might now work with very long articles or reports. Additionally, sometimes they make mistakes, generating results that do not match the information in the input text, which is known as ``hallucination''. Additionally, GPT-4 results are largely random and vary with each run, despite attempts to control its randomness through model parameter settings. We are aware of these issues and aim to elucidate that if a new and improved LLMs comes along, we can easily incorporate the model into our framework. In other words, \tool{} can adapt and get better as new LLMs are developed, thereby generating more accurate and reliable results from the text in the future.

\textbf{Numerical Data Dependency.} While our method can infer numerical data from textual context, it requires numerical information, whether stated explicitly or implied within the context. Because it is designed to interpret and extract quantitative data from text by discerning underlying values suggested by the context. In other words, Without data within text, our method cannot generate accurate charts. This highlights the importance of at least minimal numerical information within the document for our method to effectively function.




\textbf{Less Scatter Plot for Text.} When developing \tool{}, we observed that texts are typically transformed into line charts, bar charts, and pie charts, with scatter plots being less common in the transformation. This observation indicates the nature of the text. Texts that describe trends over time, such as sales figures over several months or temperature changes throughout a year, are aptly represented by line charts. Bar charts are most suitable for texts that compare quantities across different categories such as the population sizes of various cities or the revenue of different products. Pie charts are ideal for texts discussing parts of a whole, like the market share of companies within an industry, which delineates proportions in a visually intuitive manner. The rarity of scatter plots is attributed to the requirement of the numerous data. 
\alexin{However, numerous data rarely exists in the text. Thus \tool{} seldom generates scatter plots from the text.}

\textbf{Long Document Navigation Support.} The \tool{} assists users navigate long documents by extracting and organizing data related to a specific topic within the text into a chart. For example, given a comprehensive report on global climate change, a user is interested in a text about the topic of the impact of renewable energy adoption. \tool{} can extract relevant data scattered across the report, and then accurately assemble them into a chart. The chart serves as a focused lens, guiding users to locations of the text where the topic is discussed in detail. By doing this, our approach facilitates a targeted exploration of the document, allowing users to efficiently sift through vast amounts of information to find data that is more relevant to his/her interests. This capability enables the user
to quickly understand complex topics without having to comb through the entire report.







%% file: src/6-conclusion.tex
\section{Conclusion}
We proposed \tool{}, an innovative solution designed to transform complex data-involved texts into intuitive and expressive visualizations. It consists of a tabular data inference module and a chart generation module. These modules adeptly address the complexities inhere in data-involved text  - ambiguity, data sparsity, and subjective sentiment - to generate tabular data and expressive charts that convey the underlying data and insights. We conducted comprehensive evaluations, including case studies, expert interviews, and user studies, and \tool{} has proven to be useful in helping users understand insights from data-involved text documents easily. 

In the future, we plan to extend our work to include more chart types such as heatmaps, histograms, and stacked bar charts. Additionally, it will be interesting to explore the possibility of incorporating external data sources to help \tool{} infer values more accurately from data-involved texts.

%% file: main.bbl
\begin{thebibliography}{10}
\bibitem{religiously_unaffiliated}
About three-in-ten u.s. adults are now religiously unaffiliated.
\end{thebibliography}
